\documentclass[aps,prl,showpacs,twocolumn,superscriptaddress]{revtex4}
\AtBeginDvi{}
\usepackage{graphicx}
\usepackage{epsfig}
\usepackage{dcolumn}
\usepackage{bm}

\begin{document}
\title{Phase space deformation of a trapped dipolar Fermi gas}
\date{\today}

\author{Takahiko Miyakawa}
\affiliation{Department of Physics, Faculty of Science, Tokyo University of Science,
1-3 Kagurazaka, Shinjuku, Tokyo, 162-8601, Japan}
\author{Takaaki Sogo}
\affiliation{Department of Physics, Kyoto University, Kitashirakawa, Sakyo, Kyoto606-8502, Japan}
\author{Han Pu}
\affiliation{Department of Physics and Astronomy, and Rice Quantum Institute, Rice University,
Houston, Texas 77251-1892, USA}

\begin{abstract}
We consider a system of quantum degenerate spin polarized fermions
in a harmonic trap at zero temperature, interacting via
dipole-dipole forces. We introduce a variational Wigner function
to describe the deformation and compression of the Fermi gas in
phase space and use it to examine the stability of the system. We
emphasize the important roles played by the Fock exchange term of
the dipolar interaction which results in a non-spherical Fermi
surface.
\end{abstract}


\maketitle

Two-body collisions in usual ultracold atomic systems can be
described by short-range interactions. The successful realization
of chromium Bose-Einstein condensate (BEC) \cite{Griesmaier05} and
recent progress in creating heteronuclear polar molecules
\cite{Stan04} have stimulated great interest in quantum degenerate
dipolar gases. The anisotropic and long-range nature of the
dipolar interaction makes the dipolar systems different from
non-dipolar ones in many qualitative ways \cite{review}. Although
most of the theoretical studies of dipolar gases have been focused
on dipolar BECs, where the stability and excitations of the system
are investigated (see Ref.~\cite{review} and references therein,
and also Ref.~\cite{shai}) and new quantum phases are predicted
\cite{lattice,spinor}, some interesting works about dipolar Fermi
gas do exist. These studies concern the ground state properties
\cite{Goral01,Goral03,Dutta06}, dipolar-induced
superfluidity~\cite{Baranov02}, and strongly correlated states in
rotating dipolar Fermi gases~\cite{Baranov05}. None of these
studies, however, takes the Fock exchange term of dipolar
interaction into proper account \cite{noteex}.

In this Letter, we study a system of dipolar spin polarized Fermi
gas. We will show that the Fock exchange term that is neglected in
previous studies plays a crucial role. In particular, it leads to
the deformation of Fermi surface which controls the properties of
fermionic systems, and it affects the stability property of the
system. As Fermi surface can be readily imaged using time-of-flight
technique \cite{surface}, this property thus offers a
straightforward way of detecting dipolar effects in Fermi gases.

In our work, we consider a trapped dipolar gas of single component
fermions of mass $m$ and magnetic or electric dipole moment $\bm
d$ at zero temperature. The dipoles are assumed to be polarized
along the $z$-axis. The system is described by the Hamiltonian
\begin{equation}
\label{Hamiltonian}
   H=\sum_{i=1}^{N}  \left[-\frac{\hbar^2}{2m}\nabla_i^2 + U(\bm
   r_i)\right]
      + \frac{1}{2}\sum_{i \neq j} V_{dd}(\bm r_i - \bm r_j)\,,
\end{equation}
where $V_{dd}(\bm r)=(d^2/r^3)(1-3z^2/r^2)$ is the two-body
dipolar interaction and $U(\bm r)$ the trap potential.
To characterize the system, we use a semiclassical approach in
which the one-body density matrix is given by
\begin{equation}
   \rho(\bm r, \bm r^\prime)=\int \frac{d^3k}{(2\pi)^3}\, f\left(\frac{\bm r + \bm r^\prime}{2}
   ,\bm k\right) {\rm e}^{i \bm k (\bm r -\bm r^\prime)}\,,
\end{equation}
where $f(\bm r, \bm k)$ is the Wigner distribution function. The
density distributions in real and momentum space are then given
respectively by
\begin{eqnarray*}
n(\bm r) &=& \rho (\bm r, \bm r) =(2\pi)^{-3}\int {d^3k}\,
f\left(\bm r
   ,\bm k\right) \,, \\
\tilde{n} (\bm k) &=&(2\pi)^{-3}\int {d^3r}\, f\left(\bm r
   ,\bm k\right) \,.
\end{eqnarray*}

Our goal is to examine $n(\bm r)$ and $\tilde{n} (\bm k)$, as well
as the stability of the system by minimizing the energy functional
using a variational method. Within the Thomas-Fermi-Dirac
approximation~\cite{Goral01}, the total energy of the system is
given by $E = E_{kin}+E_{tr}+E_{d}+E_{ex}$, where
\begin{eqnarray}
\label{Ekinetic}
   E_{kin} &=& \int d^3r\int\frac{d^3k}{(2\pi)^3}\, \frac{\hbar^2 \bm k^2}{2m} f(\bm r,\bm k), \\
\label{Etrap}
   E_{tr} &=& \int d^3r \, U(\bm r) n(\bm r), \\
\label{EHartree}
   E_{d} &=& \frac{1}{2}\int d^3r\int d^3r^\prime\,
                         V_{dd}(\bm r - \bm r^\prime)\, n(\bm r)\,n(\bm r^\prime) , \\
\label{EFock}
   E_{ex} &=& -\frac{1}{2}\int d^3r\int d^3r^\prime \int\frac{d^3k}{(2\pi)^3}
   \int\frac{d^3k^\prime}{(2\pi)^3}\, V_{dd}(\bm r -\bm r^\prime) \nonumber \\
   &\times& {\rm e}^{i(\bm k - \bm k^\prime)\cdot (\bm r - \bm r^\prime)}
   f\left(\frac{\bm r + \bm r^\prime}{2},\bm k\right)
   f\left(\frac{\bm r + \bm r^\prime}{2},\bm k^\prime\right).
\end{eqnarray}
The dipolar interaction induces two contributions: the Hartree
direct energy $E_d$ and the Fock exchange energy $E_{ex}$. The
latter arises due to the requirement of the antisymmetrization of
many-body fermion wave functions and is therefore absent for the
dipolar BECs.

{\em Homogeneous case ---} Let us first consider a homogeneous
system of volume $\cal{V}$ with number density $n_f$, which will
provide some insights into the trapped system to be discussed
later. In this case, we obviously have $E_{tr}=0$. We choose a
variational ansatz for the Wigner distribution function that is
spatially invariant:
\begin{equation}
\label{varWigner}
   f(\bm r, \bm k) = f(\bm k)= \Theta\left( k^2_F -\frac{1}{\alpha}(k_x^2+k_y^2) -\alpha^2 k_z^2 \right),
\end{equation}
where $\Theta()$ is Heaviside's step function. Here the positive
parameter $\alpha$ represents deformation of Fermi surface~\cite{RingSchuck},
the constant $k_F$ is the Fermi wave number
and is related to the number density through $n_f=k_F^3/6\pi^2$.
The choice of (\ref{varWigner}) preserves the
number density, i.e., $(2\pi)^{-3} \int d^3k \,f(\bm k) =n_f$.

The exchange energy can be rewritten as
\begin{eqnarray*}
   E_{ex} &=& - \frac{\cal{V}}{2} \int \frac{d^3k}{(2\pi)^3} \int \frac{d^3k^\prime}{(2\pi)^3}\,
   f(\bm k) f(\bm k^\prime) \tilde{V}_{dd}(\bm k - \bm k^\prime)\,,
\end{eqnarray*}
Here we have used the Fourier transform of the dipolar potential
$\tilde{V}_{dd}(\bm q)=(4\pi/3)d^2(3\cos^2 \theta_{\bm q}-1)$
where $\theta_{\bm q}$ is the angle between the momentum $\bm q$
and the dipolar direction (i.e., the $z$-axis)~\cite{Goral02}. We
note that the Hartree direct term becomes zero for uniform density
distribution of fermions because the average over the angle
$\theta_{\bm q}$ cancels out the interaction effect.

\begin{figure}
    \begin{center}
    \includegraphics[width=6.cm]{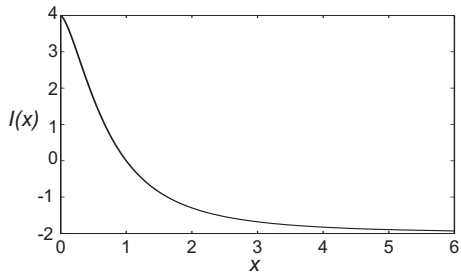}
    \end{center}
    \caption{Deformation function $I(x)$.}
    \label{func}
\end{figure}
Using the variation ansatz (\ref{varWigner}), the exchange energy
can be evaluated analytically and is given by
\begin{equation}
\label{exho}
   E_{ex}=-\frac{\pi d^2 {\cal V}}{3}\,I(\alpha)\,n_f^2\,,
\end{equation}
where we have defined the ``deformation function":
\begin{eqnarray*}
I(x)=\int_0^\pi d\theta \, \sin \theta \, \left( \frac{3\cos^2
\theta}{x^3 \sin^2 \theta + \cos^2 \theta}-1 \right) \,.
\end{eqnarray*}
This integral has rather complicated analytical form. It is more
instructive to plot out the function $I(x)$ which we show in
Fig.~\ref{func}. $I(x)$ is a monotonically decreasing function of
$x$, positive for $x<1$, passing through zero at $x=1$ and becomes
negative for $x>1$. The exchange energy (\ref{exho}) therefore
tends to stretch the Fermi surface along the $z$-axis by taking
$\alpha \rightarrow 0$. This however comes with the expense of the
kinetic energy
\[E_{kin}=\frac{\cal V}{5}\,\frac{\hbar^2 k_F^2}{2m}\, n_f
\left(\frac{1}{\alpha^2}+2\alpha\right) \,,\] which favors an
isotropic spherical Fermi surface (i.e., $\alpha=1$). The
competition between the two will find an optimal value of $\alpha$
in the region $\alpha \in (0,\,1)$. The dipolar interaction
therefore, through the Fock exchange energy, deforms the Fermi
surface of the system. This may be regarded as the
magnetostriction effect in momentum space.

{\em Inhomogeneous case ---} Let us now turn to a system of $N$
atoms confined in a harmonic trapping potential with axial symmetry:
\[U(\bm r) = \frac{1}{2}m [\omega_r^2\,(x^2+y^2)+\omega_z^2 \,z^2]
\,.\] We choose a variational Wigner function that has the same form
as in the homogeneous case, i.e., Eq.~(\ref{varWigner}), but now the
Fermi wave number $k_F$ is no longer a constant and has the
following spatial dependence:
\begin{equation}
   k_F(\bm r) = \left\{\tilde{k}_F^2 - \frac{\lambda^2}{a_{ho}^4}\left[ \beta (x^2+y^2) + \frac{1}{\beta^2}z^2
   \right]
                         \right\}^{1/2},
\end{equation}
where $a_{ho}=\sqrt{\hbar/m\omega}$ and
$\omega=(\omega_r^2\omega_z)^{1/3}$. The variational parameters
$\beta$ and $\lambda$ represent deformation and compression of the
dipolar gas in real space, respectively. Using $N=\int d^3r\,
n(\bm r)$ one can easily find that $\tilde{k}_F =
(48N)^{1/6}\lambda^{1/2}/a_{ho}$. The corresponding density
distributions in real and momentum space are give by
\begin{eqnarray*}
   n(\bm r) &=& \frac{\tilde{k}^3_F}{6\pi^2}\left\{1 -
   \frac{1}{R_F^2}\left[
                  \beta(x^2+y^2)+\frac{1}{\beta^2}
                  z^2\right] \right\}^{3/2}\,, \\
 \tilde{n}(\bm k) &=& \frac{R^3_F}{6\pi^2}\left\{1 - \frac{1}{\tilde{k}_F^2}
                   \left[\frac{1}{\alpha}\left({k^2_x}
                     +k_y^2\right)-\alpha^2 {k_z^2} \right]\right\}^{3/2}\,,
\end{eqnarray*}
respectively, where $R_F= (48N)^{1/6}a_{ho}/\lambda^{1/2}$.

Under this ansatz, each term in the energy functional can be
evaluated analytically, with the total energy given by, in units of
$N^{4/3}\hbar \omega$,
\begin{eqnarray}
   \epsilon(\alpha,\beta,\lambda)&=&c_1\left[\lambda\left(2\alpha+\frac{1}{\alpha^2}\right)
   +\frac{1}{\lambda}\left(2\frac{\beta_0}{\beta}+\frac{\beta^2}{\beta_0^2}\right)\right]\nonumber\\
 \label{varenergy}
   &+&c_2 c_{dd} N^{1/6}\lambda^{3/2} \{I(\beta)-I(\alpha)\}\,,
\end{eqnarray}
where $c_1=3^{1/3}/2^{8/3}\simeq 0.2271$, $c_2=2^{10}/(3^{7/2}\cdot
5 \cdot 7 \pi^2)\simeq 0.0634$, $c_{dd}=d^2/(a_{ho}^3 \hbar\omega)$, and $\beta_0
\equiv (\omega_r/\omega_z)^{2/3}$ measures the trap aspect ratio.
Here the two terms in the square bracket at rhs represent the
kinetic and trapping energy, respectively, while those in the curly
bracket are the direct and exchange interaction terms, respectively.

It is not difficult to see that Eq.~(\ref{varenergy}) is not
bounded from below, a result arising from the fact that the
dipolar interaction is partially attractive. There however exists,
under certain conditions, a local minimum in (\ref{varenergy}),
representing a metastable state. This situation is reminiscent of
the case of a trapped attractive BEC \cite{abec}. For the
metastable state, the variational energy
$\epsilon(\alpha,\beta,\lambda)$ satisfies the Virial theorem
$2E_{kin}-2E_{tr}+3(E_{d}+E_{ex})=0$. Hereafter, we refer to the
metastable state as the ground state.

\begin{figure}
   \begin{center}
    \includegraphics[width=9.5cm]{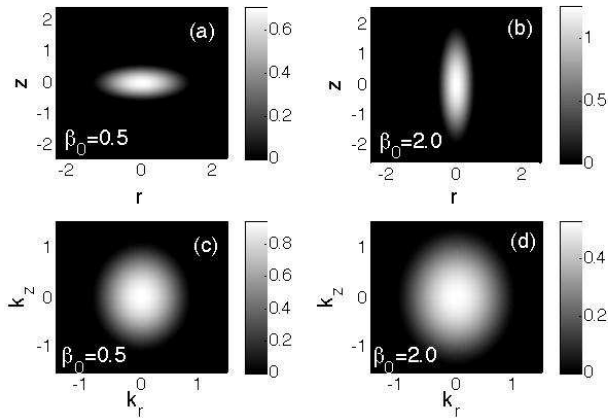}
   \end{center}
    \caption{Density distributions in real space (upper plots, in units of $\tilde{k}_F^3$)
    and in momentum space (lower plots, in units of $R_F^3$)
    for an oblate trap with $\beta_0=0.5$ (left plots) and
    a prolate trap with $\beta_0=2$ (right plots). Here $r=\sqrt{x^2+y^2}$, $k_r=\sqrt{k_x^2+
    k_y^2}$ and $c_{dd}N^{1/6}=1.5$.}
    \label{grey}
\end{figure}

We find the ground state by numerically minimizing
Eq.~(\ref{varenergy}). Fig.~\ref{grey} shows the ground state
density distributions in both real and momentum space for two
different traps. One can see that while the spatial density
distributions are essentially determined by the trap geometry, the
momentum density distributions by contrast are quite insensitive
to the trapping potential and are in both cases elongated along
the dipolar direction. Further, the momentum central density at
$\bm k=0$ decreases as $\beta_0$ increases.

\begin{figure}
    \begin{center}
    \includegraphics[width=7.5cm]{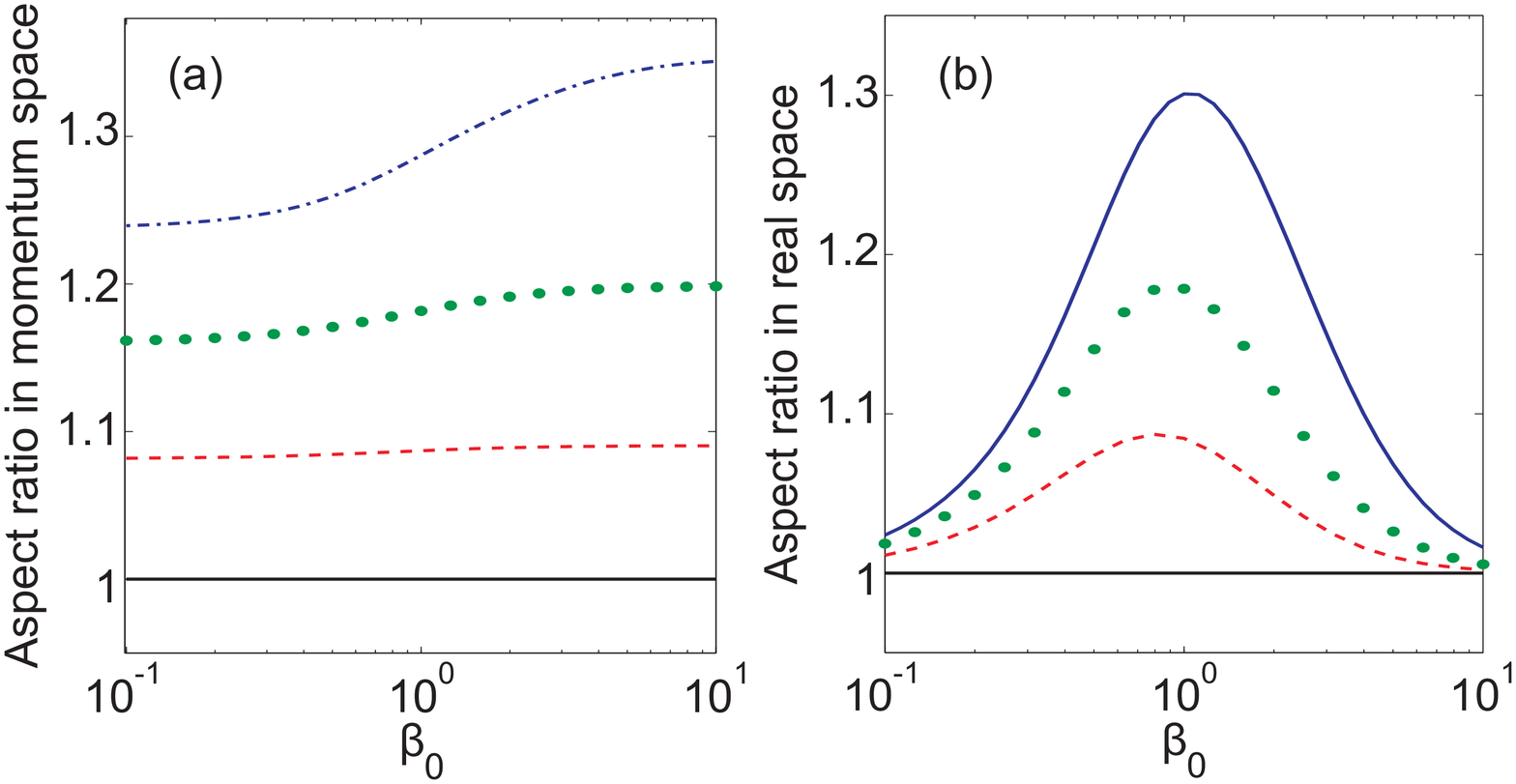}
    \end{center}
    \caption{(Color online) (a) Aspect ratio in momentum space
    $\sqrt{\langle k_z^2 \rangle_0/\langle k_x^2 \rangle_0}$ and (b)
    aspect ratio in real space $\sqrt{\langle z^2 \rangle_0/\langle x^2
    \rangle_0}$ normalized to that for a non-interacting system
    as functions of $\beta_0$
    for $c_{dd}N^{1/6}=0$ (solid line), $0.5$ (dashed line),
    $1$ (dotted line), and $1.5$ (dot-dashed line).}
    \label{momdensity}
\end{figure}

The stretch in $k_z$ is more clearly illustrated in
Fig.~\ref{momdensity}(a), where we have plotted the ratio of the
root mean square momentum in $k_z$ direction, $\sqrt{\langle k^2_z
\rangle_0}$, and to that in $k_x$ direction, $\sqrt{\langle k^2_x
\rangle_0}$, as a function of the trap aspect ratio $\beta_0$ for
several dipolar strengths. It turns out that the dipolar
interaction leads to nonspherical momentum distribution stretched
along the dipolar direction irrespective to the geometry of
trapping potential. This can be attributed to the Fock exchange
energy that becomes negative for $\alpha<1$ as discussed in the
homogeneous system. This result is in stark contrast to the case
of dipolar BEC in which the Fock exchange energy is absent and the shape of
the momentum distribution is related to that of the spatial
distribution through the Fourier transformation. Note that, for
non-interacting fermions, the resulting momentum distribution is
isotropic independent of the trapping
potential~\cite{Vichi98,PSBook}.

\begin{figure}
    \begin{center}
    \includegraphics[width=6.8cm]{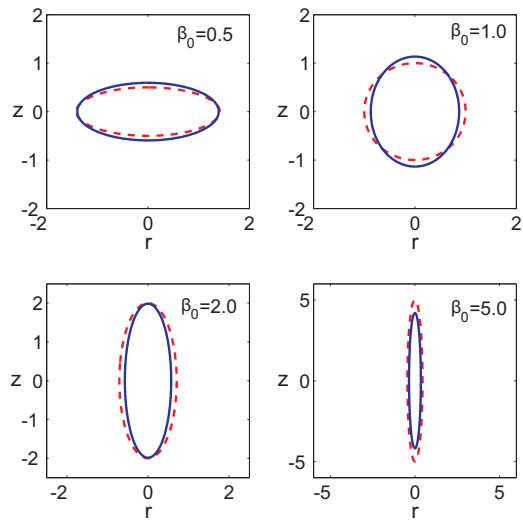}
    \end{center}
    \caption{(Color online) Thomas-Fermi surface in real space of the ground state for noninteracting case (dashed line)
    and  for interacting case $c_{dd}N^{1/6}=1.5$ (solid line).}
    \label{Rsurface}
\end{figure}

Figure~\ref{Rsurface} shows the real space Thomas-Fermi surface of
the ground state for different trap geometry. The surface of
non-interacting fermions has been plotted as dashed lines for
comparison. While the shape of the cloud relies on the trap
geometry, the dipolar interaction tends to stretch the gas along
the dipolar direction also in real space while compress the gas
along the perpendicular radial direction. However, once the
trapping potential becomes highly elongated (i.e., $\beta_0 \gg
1$), the dipolar interaction tends to shrink the whole cloud in
both the radial and the axial directions as shown in the case for
$\beta_0=5$ in Fig.~\ref{Rsurface}. This is because, for such a
cigar-shaped trap, a number of dipolar fermions align in the axial
(dipolar) direction and feel strong mutual attractions.

To better quantify the real space deformation, we show in
Fig.~\ref{momdensity}(b) the aspect ratio of the cloud
$\sqrt{\langle z^2 \rangle_0/\langle x^2 \rangle_0}$, normalized
to that of non-interacting Fermi gas, for different trap geometry.
The deviation of the aspect ratio from non-interacting case is
most dramatic for near spherical traps with $\beta_0 \approx 1$.

\begin{figure}
    \begin{center}
    \includegraphics[width=6.cm]{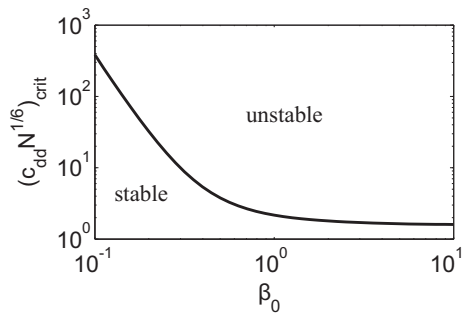}
    \end{center}
    \caption{Critical dipolar interaction strength $c_{dd}N^{1/6}$ as a function of
    $\beta_0$.}
    \label{critical}
\end{figure}

As we have already pointed out, only metastable dipolar gas can
exist in a trap. For sufficiently strong dipolar interaction
strengths, Eq.~(\ref{varenergy}) no longer supports the local
minimum and the system is expected to be unstable against
collapse. For a given trap aspect ratio $\beta_0$, the onset of
this instability gives rise to a critical value for
$c_{dd}N^{1/6}$, yielding the phase diagram as shown in
Fig.~\ref{critical}. In Ref.~\cite{Goral01}, G\'{o}ral {\it et
al.} showed that, for a sufficiently oblate trap, the system is
always stable regardless of the strength of the dipolar
interaction. The critical trap aspect ratio they found is, when
translating to our notation, about $\beta_0 \approx 0.33$. Our
calculation, however, indicates that no such critical value exists
in $\beta_0$
--- when the trap becomes more and more oblate, the
critical dipolar strength increases rapidly but always remains
finite. This contradiction between our work and theirs originates
from the Fock exchange term of the dipolar interaction that plays
a significant role near critical point and is not properly treated
in Ref.~\cite{Goral01}.

Finally, let us briefly discuss how to detect the dipolar effects
in Fermi gases. In fact, the momentum space magnetostriction
effect indicates that the dipolar effects will manifest in
time-of-flight images. Assuming ballistic expansion after turning
off the trapping potential, for time $t \gg 1/\omega$, the aspect
ratio of the expanded cloud approaches the initial in-trap aspect
ratio of the momentum distribution as shown in
Fig.~\ref{momdensity}~(a). In other words, regardless of the
initial trap geometry, the expanded cloud will eventually become
elongated in the dipolar direction. In comparison, the expansion
of a non-interacting Fermi gas is isotropic and results in a
spherical cloud in the long time limit \cite{Vichi98,PSBook}. The
expansion technique has also been used to detect the dipolar
effects in chromium BEC. In that case, the expansion dynamics
strongly depends on the initial trap geometry
\cite{Giovanazzi06,Lahaye07}.

In conclusion, we have analyzed the properties of a dipolar Fermi
gas using a variational method. We have emphasized the important
roles played by the Fock exchange energy. We found that the
dipolar interaction induces deformation of the Fermi surface and
of the phase space density distribution. The resulting anisotropic
momentum distribution of the dipolar gas, which can be readily
probed using time-of-flight technique, is elongated in dipolar
direction irrespective to the trap geometry.

Future work will extend these considerations into collective
excitations and superfluidity of the dipolar gas. Since Fermi
surface is a key ingredient for low energy properties of Fermi
gases, a non-spherical Fermi surface will lead to new features in
collective phenomena in dipolar Fermi gases.

This work is supported by the Grant-in-Aid for the 21st Century COE ``Center for Diversity and
Universality in Physics" from the Ministry of Education,
Culture, Sports, Science and Technology (MEXT) of Japan.
HP acknowledges support from NSF, the Robert A. Welch Foundation
and the W. M. Keck Foundation.



\begin{references}
\bibitem{noteex} The exchange term is considered in Ref. \cite{Goral01}.
However, without taking into account the possibility of the
deformation of Fermi surface, their treatment of the exchange term
is not appropriate.
\end{references}
\end{document}